\documentstyle[11pt,aaspp]{article}
\input psfig
\psfull
\singlespace

\def\be{\begin{equation}}
\def\ee{\end{equation}}
\def\ba{\begin{eqnarray}}
\def\ea{\end{eqnarray}}
\def\etal   {{\sl et al.}~\rm}
\def\apj    {{\sl Ap.J.$\!$}~\rm}
\def\deg    {^\circ\!}
\newcommand {\gtsim}{\mbox{$\stackrel{>}{_{\sim}}$} }

\begin{document}
\title{The Dipole Observed in the {\it COBE~} DMR Four-Year Data}
\author{C. H. Lineweaver\altaffilmark{1,2},
L. Tenorio\altaffilmark{3},
G. F. Smoot\altaffilmark{4},
P. Keegstra\altaffilmark{5},
A. J. Banday\altaffilmark{5,6} \&
P. Lubin\altaffilmark{7}
}
\altaffiltext{1}{Observatoire de Strasbourg, 67000 Strasbourg, France.}
\altaffiltext{2}{e-mail: charley@cdsxb6.u-strasbg.fr}
\altaffiltext{3}{Universidad Carlos III, Madrid, Spain.}
\altaffiltext{4}{LBL, SSL \& CfPA, Building 50-205, University of California,
Berkeley CA 94720.}
\altaffiltext{5}{Hughes STX Corporation, LASP, Code 685, NASA/GSFC, Greenbelt MD 20771.}
\altaffiltext{6}{Max Planck Institut f\"ur Astrophysik, 85740, Garching, Germany.}
\altaffiltext{7}{UCSB Physics Department, Santa Barbara, CA 93106.}

\normalsize
\begin{abstract}
The largest anisotropy in the cosmic microwave background (CMB)
is the $\approx 3$ mK dipole assumed to be due to our velocity with respect 
to the CMB.
Using the four year data set from all six channels of the COBE Differential 
Microwave Radiometers (DMR), we obtain a best-fit dipole amplitude 
$3.358 \pm 0.001 \pm 0.023$ mK 
in the direction 
($\ell, b)=(264\deg.31 \pm 0\deg.04 \pm 0\deg.16, 
\:+48\deg.05 \pm 0\deg.02 \pm 0\deg.09)$,
 where the first uncertainties are statistical and the second include calibration and 
combined systematic uncertainties.
This measurement is consistent with previous DMR and FIRAS results.
\end{abstract}

\keywords{cosmic microwave background --- cosmology: observations}
\section{Introduction}
\label{sec:intro}
The Sun's motion with respect to the cosmic microwave background (CMB)
is believed to be responsible for the largest anisotropy seen in the 
COBE DMR maps: the $\approx 3$ mK dipole in the direction of the 
constellation Leo.  A measurement of this Doppler dipole 
thus tells us our velocity with respect to the rest frame of the CMB.
A high precision measurement of the dipole will be used as the primary
calibrator for an increasing number of ground, balloon and satellite 
anisotropy experiments (Bersanelli \etal 1996). 
The accurate removal of the Doppler dipole and Doppler quadrupole from
anisotropy maps improves the precision of the anisotropy results. 
The CMB dipole is also used to calibrate 
bulk flow observations which 
yield independent but much less precise dipole values.
In addition, anisotropy measurements in other background 
radiations will be made in the future and an eventual test of the 
Doppler origin of the CMB dipole  will be facilitated by a CMB dipole of 
maximum precision (Lineweaver \etal 1995).
In this paper we use the DMR four-year data to determine the precise direction 
and the amplitude of the observed dipole.
The largest source of directional error (aliasing of CMB power combined with
instrument noise) has been reduced by using relatively 
small Galactic plane cuts.

In Section 2
we discuss the data analysis and in Section 3
we discuss contamination from Galactic emission as well as other factors 
contributing to the error budget. 
In Section 4 we present our results. 
We then discuss and compare our results to FIRAS and 
other reported DMR dipole results.

\section{Data Analysis}
\label{sec:analysis}

The four year DMR data set and its systematic errors and calibration
procedures are described in Kogut \etal (1996b).
There are 2 independent channels at each of the 3 frequencies 31.5, 53 
and 90 GHz.
We base our results on all six DMR channels since 
the less sensitive 31 GHz channels provide useful information on the
frequency dependence of Galactic contamination.

We use three methods to obtain the CMB dipole amplitude and direction.
These methods differ in the form of the input data but all of them are 
least-squares fits of the data to the coefficients of a 
spherical harmonic decomposition of the sky: 
$T(\theta, \phi)=\sum_{\ell m} a_{\ell m}Y_{\ell m}(\theta,\phi)$, where
the $Y_{\ell m}$ are real-valued spherical harmonics as described in
Smoot \etal (1991) and the dipole vector is
$\vec{D}=\sqrt{3/4\pi}(-a_{1,1},-a_{-1,1},a_{1,0})$.
To obtain the dipole we minimize the three quantities:
\ba
\label{eq:dhi}
\sum_{i} &\{ T_{i}            &- \sum_{\ell=0}^{\ell_{max}}\sum_{m=-\ell}^{+\ell} a_{\ell m}  Y_{\ell m}(i)\;\}^2/\sigma_{i}^{2}\\
\label{eq:devlong}
\sum_{t} &\{ \Delta T(t)      &- \sum_{\ell=0}^{\ell_{max}}\sum_{m=-\ell}^{+\ell} a_{\ell m} [Y_{\ell m}(t_{+})-Y_{\ell m}(t_{-})]\;\}^{2}/\sigma_{t}^{2}\\
\label{eq:dmp}
\sum_{i,j>i}&\{ \Delta T_{ij} &-\sum_{\ell=0}^{\ell_{max}}\sum_{m=-\ell}^{+\ell} a_{\ell m} [Y_{\ell m}(i)-Y_{\ell m}(j)]\;\}^{2}/\sigma_{ij}^{2}
\ea
where $T_{i}$ is a pixelized DMR temperature map,
$\Delta T(t)$ is a single DMR differential measurement
and  $Y_{\ell m}(t_{+})$ and $Y_{\ell m}(t_{-})$ are the spherical harmonics 
evaluated in the pointing directions of the DMR ``$+$'' and ``$-$'' horns
respectively, at time $t$.
The pixel-pair data, $\Delta T_{ij}$, is the average over all single 
measurements $\Delta T(t)$ where the antennas are pointing at 
pixels $i$ and $j$.  The denominators are the variances of the input data.
Thus, with method 1, the sum is over all map pixels, with method 2, 
the sum is over all the time-ordered data (with no pixelization)
and in method 3 the sum is over all pixel-pairs.
The three methods are consistent and agree to within the relatively small noise-only error bars
for each channel.
We use the difference between the non-pixelized method and the mean of the
two pixelized methods to estimate and correct for the smoothing due to
data pixelization.
We adopt the mean of these three methods and include
the difference in the combined systematic uncertainty.
We correct for beam smoothing by multiplying the amplitude by the factor 
1.005 (Wright \etal 1994).

\section{Analysis of Galactic Plane Cuts} 
\label{sec:errors}
\subsection{Galactic Contamination}
We estimate the influence of Galactic emission on the measurement
by solving for the dipoles in equations (\ref{eq:dhi}), (\ref{eq:devlong}), and
(\ref{eq:dmp}) for a series of Galactic plane latitude cuts.
The dipole amplitude and direction results from 
each channel and each Galactic plane cut are shown in 
Figure 1.
Galactic emission produces a dipole which pulls the solutions towards it. 
This is easily seen in Figure 1 from the locations of the $0\deg$ and $5\deg$ cut 
solutions relative to the cluster of higher cut results on the right.
Since the Galactic dipole vector is nearly orthogonal to the CMB dipole
vector, it is almost maximally effective in influencing the CMB dipole 
direction and almost minimally effective in influencing the CMB dipole 
amplitude.
We can get a rough estimate of the Galactic
dipoles by noting that the $0\deg$ cut solutions for 31, 53 and 90 
GHz are displaced from the direction of our best-fit CMB dipole
by angles $\alpha_{\nu} \approx (5\deg,\; 1\deg,\; 0\deg.5)$ respectively.
Thus the ratios of the Galactic dipoles to the CMB dipole are 
$\frac{D_{Gal,\nu}}{D_{CMB}} \approx sin(\alpha_{\nu}) 
\approx (9\%, 2\%, 1\%)$.
In 
Figure 1, 
the general increase of the dipole amplitudes seen in the 
top panel as the Galactic cut increases from  $0\deg$ to $5\deg$ 
to $10\deg$ can be explained by the fact that the Galactic dipole 
vector contains a component in the direction opposite to the CMB dipole 
(the Galactic center is $\approx 94\deg$ away) and thus reduces the total dipole in the maps.
A rough estimate of this effect on the dipole solutions D is in good 
agreement with the plot: 
 $\Delta D_{\nu} \sim sin(4\deg\;)\;sin(\alpha_{\nu})D \sim (20\,\mu$K, $5\,\mu$K, $2\,\mu$K)
for 31, 53 and 90 GHz respectively. 

Figure 1
clearly shows the influence of the Galaxy for
the $0\deg$ and $5\deg$ cuts as well as the relative agreement of the
independent channel results for both amplitude and direction.  It is
also apparent that to first approximation a $10\deg$ cut is sufficient
to remove the effect of the Galaxy on the direction of the best-fit
dipole; increases of the cut from $10\deg$ to $15\deg$ and so on, do
not push the directions away from the Galactic center or in any other
particular direction.  The results tend to cluster together.  The
directional precision of the various channels and Galactic cuts is
seen to be $\sim 0\deg.3$ and it is perhaps reassuring to note that at
the bottom and the top of the cluster are the least sensitive 31A and
31B solutions.

Figure 2
minimizes the confusion of taking a closer look at 
the cluster of points in Figure 1.
 It analyzes the directional changes 
of the dipoles in the bottom panel of Figure 1.
For example  consider the 31A results.
The angular difference between the $5\deg$ and $10\deg$ cut solutions is
a vector of length $\approx 1\deg$ starting from the $5\deg$ cut on the left and extending to the $10\deg$
cut on the right. Averaging this vector with the analogous vector from 31B,
we obtain the long thin line that runs across most of the lower panel in
Figure 2. The size of this angular deviation ($\approx 1\deg$ ), is plotted as the triangular point
in the 5-10 bin of the top panel.
An analogous procedure was followed for all channels and Galactic cuts.
Figure 2 is thus a spectral analysis of the angular deviations from one
Galactic plane cut to another.

Galactic emission significant enough to affect the
dipole results will tend to pull the three channels in approximately 
the same direction and favor a spectral
behavior typical of synchrotron or free-free emission.
In the top panel, the two reference lines originating on the 31 GHz point 
in the 0-5 and 5-10 bins indicate this expected spectral behavior for
 synchrotron radiation (thin) and free-free emission (thick).
The results in the 0-5 and 5-10 bins are obviously from Galactic emission.
The directions in these bins are also strongly correlated.
The absence of this spectral and directional behavior for the bins 
10-15 and larger is evidence that the Galaxy is no longer the major contributor to the
directional uncertainty of the dipole.
Although the 20-25 bin seems indicative of the spectral 
behavior of Galactic emission, the incoherent directional behavior is inconsistent with a 
common spatial origin for the supposed source.

Evidence supporting the idea that Galactic emission is relatively unimportant 
is provided by the small differences between the dipole solutions using 
the  ``custom'' cut (Fig. 1 of Kogut \etal 1996a)
and the straight $|b| > 20\deg$ cut presented here.
The differences in amplitude, longitude and latitude are less than
$0.2\%, 9\%$, and $6\%$ of our error bars on these respective quantities.
If plotted in Figure 1, the ``custom'' cut solutions overlap the 
$|b| > 20\deg$ points with a barely distinguishable displacement in the 
direction of the $|b| > 25\deg$ solutions.

\subsection{Higher Multipole CMB Contamination}
For the purposes of determining the dipole there are two sources of noise;
instrument noise with a power law spectral index $n\approx 3$
and the $n\approx 1$ CMB signal.
At $10\deg$ scales the CMB signal to noise ratio in the maps is  $\sim 2$ 
(Bennett \etal 1996). Thus on larger scales the CMB signal dominates the 
instrument noise and correspondingly, the uncertainties on the dipole from 
the CMB signal are larger than those from the instrument noise.
The uncertainties from {\it both} are reduced by 
lowering the Galactic plane cut.
In the 15-20 bin of Figure 2 (and to a lesser extent in the 10-15 and 
25-30 bins) we see a directional and spectral behavior consistent with a 
common spatial origin and a CMB spectra (no frequency dependence of the 
angular deviation).
This suggests that large scale power of the CMB signal is responsible for
these displacements (rather than Galactic emission), and that a smaller 
(not a larger) cut is called for.
This is further supported by the fact that for $|b|\,\gtsim\,20\deg$, the combined
free-free and dust emission from the Galaxy at 53 and 90 GHz produces only 
$\sim 10\,\mu$K rms (Kogut \etal 1996a) while the CMB signal rms 
is $\sim 35\,\mu$K (Banday \etal 1996). 

To estimate the uncertainty in the dipole results due to the CMB signal
we simulate $n=1.2$, $Q_{rms-PS}=15.3\, \mu$K CMB skies for $2 \le \ell \le 25$.
We superimpose these maps on a known dipole and solve for the dipole using
a $15\deg$ Galactic plane cut.
No bias is detected and the rms's of the results around the input
values are $3.3\,\mu$K in amplitude, $0\deg.127$ in longitude and 
$0\deg.062$ in latitude. We include these uncertainties in the
combined systematics.

Galactic cuts greater than $15\deg$ are not useful corrections which eliminate more and more 
Galactic contamination; they introduce systematic errors associated with
large Galactic cuts due to the increasingly non-orthogonal basis 
functions $Y_{\ell m}(\theta, \phi)$, over the increasingly limited and thus noisier 
input data.
For example, simulations with a $25\deg$ cut yield rms uncertainties due to
the CMB signal $\sim 75\%$ larger than the $15\deg$  simulations.
We conclude that, for the method used here, the Galactic cuts of $10\deg$ and $15\deg$
are the best compromise to minimize the combined effect of CMB aliasing, 
Galactic contamination and noise.
The high precision of our dipole direction results depend on this conclusion.
Note that this choice for the optimal Galactic cut is smaller than the
$\approx 20\deg$ cut used when one is trying to compute the correlation function or
determine the $\ell \geq 2$ components of the power spectrum of the CMB signal
which are  smaller than the dipole by a factor of $\sim 200$. 
For such determinations, the similar compromise for simultaneously minimizing Galactic 
contamination, instrument noise and other procedural/systematic effects demands
a larger cut.

Our results are averages of the $10\deg$ and $15\deg$ cuts.
We adopt the difference between these two solutions as the 
one $\sigma$ uncertainty related to this Galactic plane cut choice.
We include this uncertainty in the combined systematics, along with
the error associated with the aliasing of the CMB signal and the
method difference errors mentioned earlier. In general, CMB aliasing
is the dominant contributor to the directional combined systematics.

\section{Results}
\label{sec:results}
Table 1 lists the weighted average of the $10\deg$ and 
$15\deg$ Galactic cut results and the uncertainties 
for each channel. 
Taking the weighted average of all six channels 
we obtain a best-fit dipole amplitude
$3.358 \pm 0.001 \pm 0.023$ mK 
in the direction 
($\ell, b)=(264\deg.31 \pm 0\deg.04 \pm 0\deg.16, 
\:+48\deg.05 \pm 0\deg.02 \pm 0\deg.09)$,
where the first uncertainties are statistical and the second are estimations
of the combined systematics.
In celestial coordinates the direction is
$(\alpha, \delta)= (11^{h}\:11^{m}\:57^{s} \pm 23^{s}, 
-7\deg.22 \pm 0\deg.08)$ (J2000).
The uncertainty in the dipole amplitude is dominated by the absolute 
calibration of the DMR instrument (Kogut \etal 1996b).
This is easily seen in Figure 1
by comparing the large error bars on our final
result (far right) with the noise-only error bars on the channel results.
 The calibration 
uncertainty plays no role in the directional uncertainty for the same reason 
that the directions of vectors $\vec{x}$ and $a\vec{x}$
(where $a$ is any positive constant) are the same. 
The uncertainty in the direction is dominated by the combined 
systematic effects discussed above.

Under the assumption that the Doppler effect is responsible for the entire CMB dipole, 
the velocity of the Sun with respect to the rest frame of the CMB is 
$v_{\odot}=369.0 \pm 2.5$ km/s, which corresponds to the dimensionless 
velocity $\beta=v_{\odot}/c = 1.231 \pm \,0.008\, \times \, 10^{-3}$.
The associated rms Doppler quadrupole%
\footnote{\small{
$Q^{2}_{rms}=\frac{4}{15}[\frac{3}{4}Q^{2}_{1}+Q^{2}_{2}+Q^{2}_{3}+Q^{2}_{4}+Q^{2}_{5}]$
where the components are defined by 
$T_{o}\frac{\beta^{2}}{2}(2cos^{2}\theta -(2/3))=$\\
$Q_{1}(3sin^{2}b -1)/2+Q_{2}sin2b\,cos\ell+
Q_{3}sin2b\,sin\ell + Q_{4}cos^2b\,cos2\ell + Q_5 cos^2 b\,sin2\ell$, where
$T_{o}$ is the mean CMB temperature and $\theta$ is the angle between the 
dipole direction and the direction of observation: $(\ell, b)$.
}}
is $Q_{rms}=1.23 \pm 0.02\; \mu$K
with components 
$[Q_{1},Q_{2},Q_{3},Q_{4},Q_{5}]= 
[0.91\pm 0.02, -0.20 \pm 0.01, -2.05\pm 0.03, -0.91\pm 0.02, 0.18\pm 0.01] \mu$K.

\section{Summary and Discussion}

We have used the DMR four-year data set to
obtain a best-fit CMB dipole amplitude
$3.358 \pm 0.023$ mK and  direction 
($\ell, b)=(264\deg.31 \pm 0\deg.17, 
\:+48\deg.05 \pm 0\deg.10)$.
Figure 3 displays the main results of this paper and 
compares them with other {\it COBE} results: 
the DMR first year (Kogut \etal 1993), 
the DMR first two years (Bennett \etal 1994),
the FIRAS (Fixsen \etal 1994, Fixsen \etal 1996) dipole results
and a pixel-based likelihood analysis of the DMR four-year 
data (Bennett \etal 1996).

Although the results are consistent,
our independent analysis differs from the Bennett \etal (1996) 
analysis in many detailed ways. 
The most important difference is our strategy for removing 
the Galactic foreground; 
we have examined the dipole results as a function of Galactic plane
cut and frequency and find that Galactic contamination of the dipole
is not important for Galactic cuts as low as $10\deg$ or $15\deg$.
The largest source of directional error, aliasing of
CMB power combined with instrument noise, can be reduced by 
using these smaller Galactic plane cuts. The result is substantially smaller
errors on the dipole direction. 

The good agreement of the DMR and FIRAS dipole results is further evidence 
that the systematic uncertainties of these two {\it COBE} instruments are 
fairly well understood.

We acknowledge the constructive comments of the anonymous referee.
We also gratefully acknowledge NASA for funding the COBE satellite and data 
processing and the many people responsible for the high quality of the
{\it COBE} DMR data. C.H.L. acknowledges
support from the French Minist\`ere des Affaires Etrang\`eres. L.T. was
partially supported by grant DGICYT PB94-0364. 

\clearpage
\footnotesize
\begin{table}
\begin{center}
Table I:  {\bf Channel Dipole Results}\\
\begin{tabular}{l r r r}  \hline
\multicolumn{1}{c}{{Channel  }} &
\multicolumn{1}{c}{{ Amplitude }} &
\multicolumn{1}{c}{{ Galactic Longitude}} &
\multicolumn{1}{c}{{ Galactic Latitude}} \\[-1.0mm]
& ($\mu$K)$^{a}$ & (degrees) & (degrees)\\ 
\hline

31A                     &    &       &    \\[-1.0mm]
Mean................    &3366&264.50&47.83\\[-1.0mm]
Total Error.........    &  85&  0.22& 0.23\\[-1.0mm]
Noise...............    &   7&  0.14& 0.09\\[-1.0mm]
Calibration.........    &  84&  0.00& 0.00\\[-1.0mm]
Combined Systematics    &  10&  0.16& 0.21\\[+0.5mm]

31B                     &    &       &    \\[-1.0mm]
Mean................    &3346&264.46&48.25\\[-1.0mm]
Total Error.........    &  77&  0.30& 0.24\\[-1.0mm]
Noise...............    &   9&  0.23& 0.12\\[-1.0mm]
Calibration.........    &  76&  0.00& 0.00\\[-1.0mm]
Combined Systematics    &   4&  0.20& 0.20\\[+0.5mm]

53A                     &    &       &    \\[-1.0mm]
Mean................    &3355&264.28&48.05\\[-1.0mm]
Total Error.........    &  23&  0.18& 0.10\\[-1.0mm]
Noise...............    &   2&  0.07& 0.03\\[-1.0mm]
Calibration.........    &  23&  0.00& 0.00\\[-1.0mm]
Combined Systematics    &   4&  0.17& 0.09\\[+0.5mm]

53B                     &    &       &    \\[-1.0mm]
Mean................    &3364&264.19&48.15\\[-1.0mm]
Total Error.........    &  24&  0.17& 0.08\\[-1.0mm]
Noise...............    &   2&  0.08& 0.04\\[-1.0mm]
Calibration.........    &  23&  0.00& 0.00\\[-1.0mm]
Combined Systematics    &   4&  0.15& 0.07\\[+0.5mm]

90A                     &    &       &    \\[-1.0mm]
Mean................    &3362&264.35&47.96\\[-1.0mm]
Total Error.........    &  67&  0.20& 0.11\\[-1.0mm]
Noise...............    &   4&  0.13& 0.07\\[-1.0mm]
Calibration.........    &  67&  0.00& 0.00\\[-1.0mm]
Combined Systematics    &   3&  0.15& 0.09\\[+0.5mm]

90B                     &    &       &    \\[-1.0mm]
Mean................    &3351&264.26&47.99\\[-1.0mm]
Total Error.........    &  43&  0.17& 0.09\\[-1.0mm]
Noise...............    &   3&  0.10& 0.05\\[-1.0mm]
Calibration.........    &  43&  0.00& 0.00\\[-1.0mm]
Combined Systematics    &   3&  0.14& 0.07\\[+0.5mm]

Total                   &      &     &    \\[-1.0mm]
Mean................    &3358&264.31&48.05\\[-1.0mm]
Total Error.........    &  23&  0.17& 0.10\\[-1.0mm]
Noise...............    &   1&  0.04& 0.02\\[-1.0mm]
Calibration.........    &  23&  0.00& 0.00\\[-1.0mm]
Combined Systematics    &   4&  0.16& 0.09\\[+0.5mm]
\hline
\end{tabular}\\
\end{center}

${}^{a}$
Values are in thermodynamic temperature transformed from antenna temperature by\\
$\Delta T = \Delta T_{ant} (e^{x}-1)^{2}/x^{2}e^{x}$ where $x= h\nu/kT_{o}$,
$T_{o}=2.73$ K. The conversion factors are thus 1.026, 1.074 and 1.227 
for 31.5, 53 and 90 GHz respectively.\\
${}^{b}$ see Kogut \etal 1996b, Table 2 for absolute calibration uncertainties.

\end{table}
\normalsize


\clearpage

\begin{figure}[h,p,t]
\psfig{file=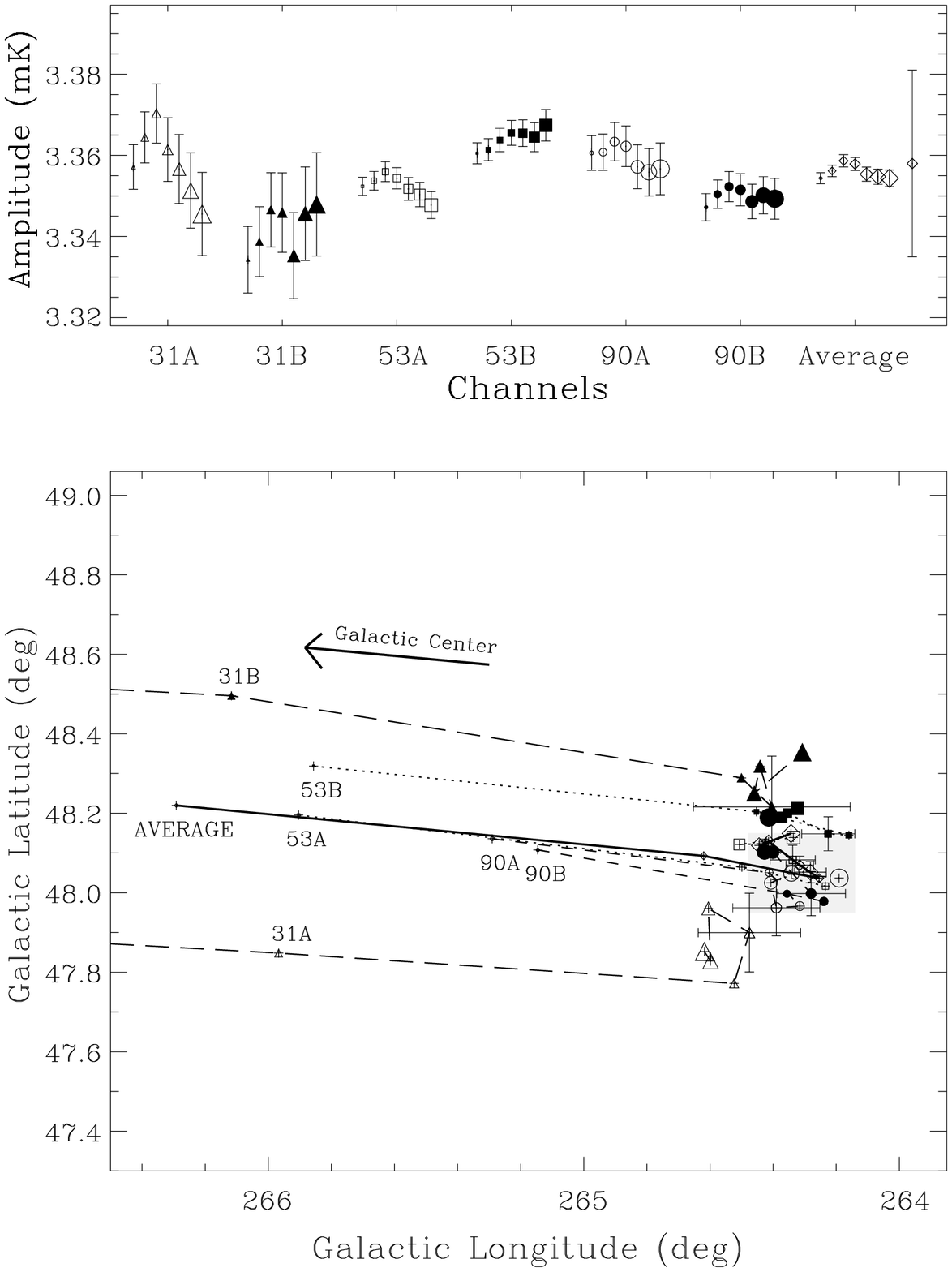,height=6.3in,width=6.0in}
\caption{Dipole amplitudes (top) and directions (bottom).
The results for each channel and
Galactic plane cut (from left to right in the top panel),
$|b| > 0\deg, \;5\deg, \;10\deg, \;15\deg, \;20\deg, \;25\deg, \;30\deg$ 
are shown.
Channels and cuts are denoted with the same point type and size in both panels.
Solutions for the dipole where no effort has been
made to eliminate Galactic emission (i.e., $0\deg$ Galactic cuts)
are labeled with the channel names 53A, 53B, 90A and 90B.
The 31 GHz labels indicate the $5\deg$ cut solutions since 
their $0\deg$ cut solutions are off the plot at longitude $\approx 271\deg$.
For each channel, the successive Galactic cuts  are connected by lines
(31: long-dashed, 53: dotted, 90: short-dashed, average: solid).
The direction of the Galactic center is toward higher
latitudes for the same reason that one flies north-west from 
London to arrive at New York.
The latitude and longitude ranges were chosen to display an approximately
square piece of the sky.
For each channel, the direction error bars on the $15\deg$ Galactic cut
solutions are shown. 
Our final dipole amplitude, including the calibration uncertainty is the point in 
the far right of the top panel. The grey box in the bottom panel denotes the
68\% confidence levels of our final dipole direction.
\label{fig.allcuts}}
\end{figure}
 

\begin{figure}[h,p,t]
\psfig{file=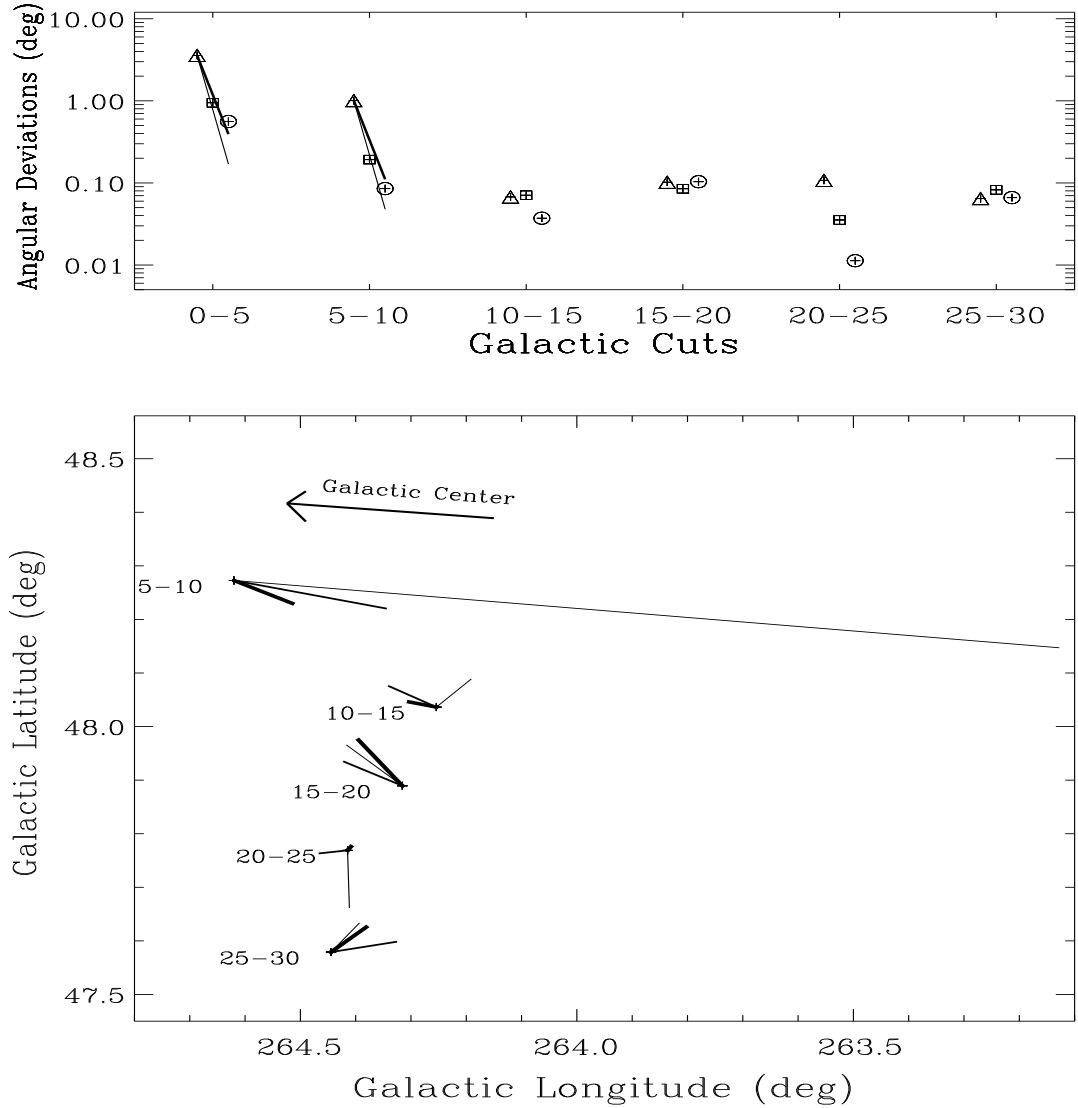,height=6.6in,width=6.0in}
\caption{Spectral analysis of the angular deviations of the dipole.
The vectors of the angular separations between the points in 
Figure 1 are indicated here in the bottom panel while 
their sizes are plotted in the top panel.
The A and B channels at each frequency, 31.5, 53 and 90,  have been 
averaged and are represented respectively by triangles, squares, circles (top)
and by thin, medium and thick lines (bottom).
In the bottom panel, for ease of comparison, the three vectors in a given bin 
originate at the same point. 
The 0-5 bin is not shown because it is similar to the
5-10 bin but (as indicated in the top panel) 
the vectors are approximately five times longer. 
In the top panel, the two reference lines originating on the 31 GHz 
point of the 0-5 and 5-10 bins indicate the expected spectral behavior if the
Galactic emission is pure synchrotron (thin) and pure free-free (thick).
The points chosen as the common origin of the vectors for each 
bin are the directions of the channel averages at the smaller of the
cuts in each bin pair.
The origin latitudes have been offset by $0.18\deg$ with respect to
each other to avoid confusion.
\label{fig.allangdevi}}
\end{figure}
\begin{figure}[h,p,t]
\psfig{file=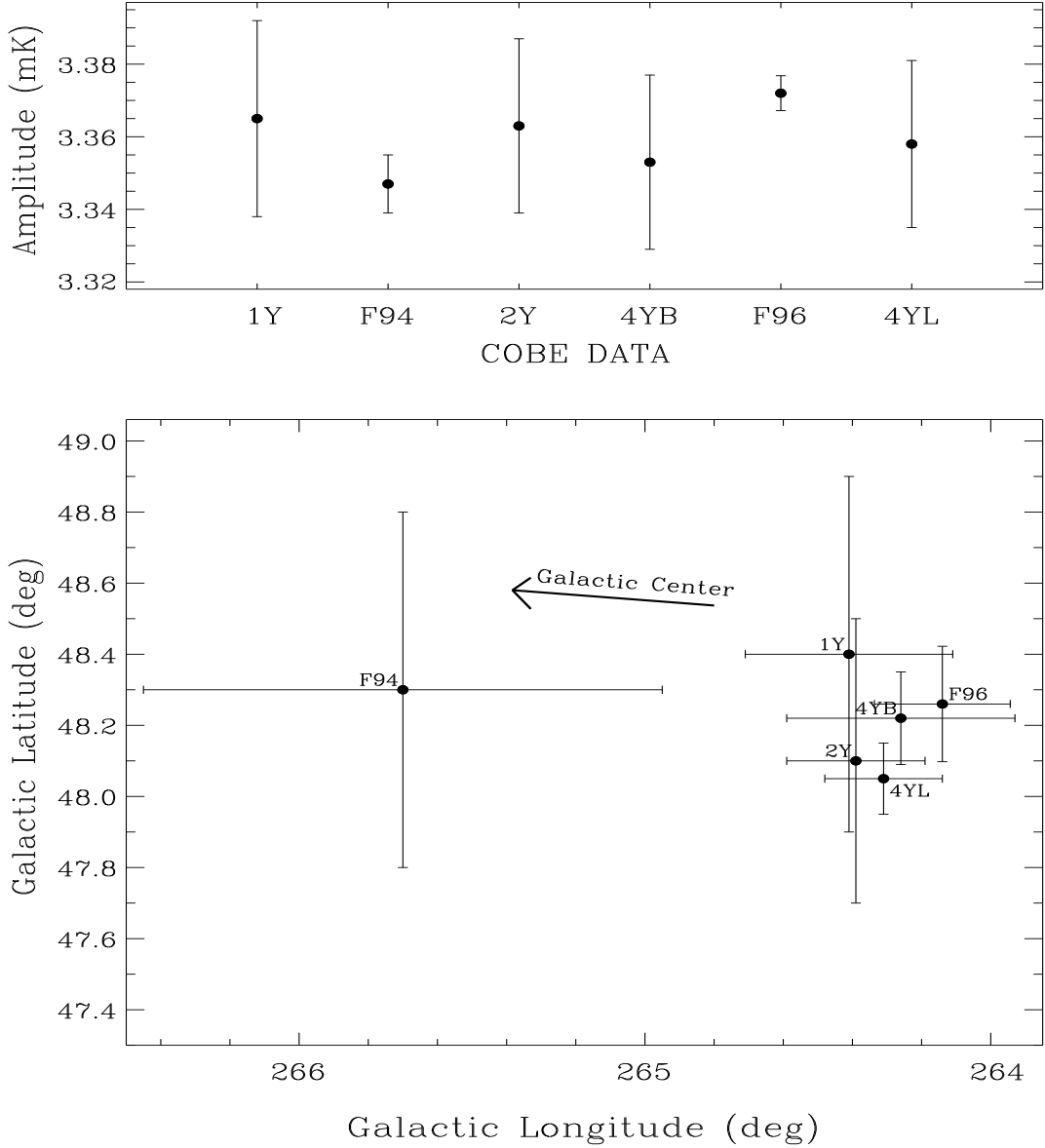,height=7.0in,width=6.0in}
\caption{  COBE dipoles.
The DMR dipole results: 1Y: (Kogut \etal 1993), 
2Y: (Bennett \etal 1994),
4YB: (Bennett \etal 1996), 
and 4YL: (this work:), are consistent with each other
and with the FIRAS results F94, F96: (Fixsen \etal 1994, 1996).
We have adjusted the published FIRAS error bars to include CMB aliasing
using the $15\deg$ cut simulation results (Section 3.2).
\label{fig.compare}}
\end{figure}


\clearpage
\begin{references}
\reference Banday, A. J., \etal 1996, \apj, in press
\reference Bennett, C. L., \etal 1994, \apj, 436, 423 
\reference Bennett, C. L., \etal 1996, \apj, in press
\reference Bersanelli, M., \etal 1996, A\&A Supp., in press
\reference Fixsen, D. J., \etal 1994, \apj, 420, 445 
\reference Fixsen, D. J., \etal 1996, \apj, submitted
\reference Kogut, A., \etal 1993, \apj, 419, 1       
\reference Kogut, A., \etal 1996a, \apj, 464, L5
\reference Kogut, A., \etal 1996b, \apj, 470, in press    
\reference Lineweaver, C. H., \etal 1995, Astrophysical Letters and Comm., 32, pp 173-181 
\reference Smoot, G. F., \etal 1991, \apj, 371, L1
\reference Wright, E. L., \etal 1994, \apj, 420, 1
\end{references}
\end{document}